\documentclass[aps,pre,twocolumn,superscriptaddress]{revtex4-1}
\usepackage{graphicx}
\usepackage{anysize}
\usepackage{lineno}
\usepackage[amssymb]{SIunits}
\usepackage{lipsum}
\usepackage{amsmath}
\usepackage{amssymb}

\newcommand \be {\begin{equation}}
\newcommand \ee {\end{equation}}
\newcommand \bea {\begin{eqnarray}}
\newcommand \eea {\end{eqnarray}}

\begin{document}

%\preprint{}

\title{Effective diffusivity of microswimmers in a crowded environment}



\author{Marvin Brun-Cosme-Bruny}
%\thanks{}
\affiliation{Univ. Grenoble Alpes, CNRS, LIPhy, F-38000 Grenoble, France}
\author{Eric Bertin}
\affiliation{Univ. Grenoble Alpes, CNRS, LIPhy, F-38000 Grenoble, France}
\author{Beno\^{\i}t Coasne}
%\thanks{}
\affiliation{Univ. Grenoble Alpes, CNRS, LIPhy, F-38000 Grenoble, France}
\author{Philippe Peyla}
%\thanks{}
\affiliation{Univ. Grenoble Alpes, CNRS, LIPhy, F-38000 Grenoble, France}
\author{Salima Rafa\"{\i}}
\email[]{salima.rafai@univ-grenoble-alpes.fr}
%\thanks{}
\affiliation{Univ. Grenoble Alpes, CNRS, LIPhy, F-38000 Grenoble, France}
%\altaffiliation{Corresponding author}


\date{\today}

%\pacs{}
%\keywords{}
\begin{abstract}
The microalga \textit{Chlamydomonas Reinhardtii} (CR) is used here as a model system to study the effect of
complex environments on the swimming of micro-organisms. Its motion can be modelled by a run and tumble mechanism so that it
describes a persistent random walk from which we can extract an effective diffusion coefficient for the large-time dynamics. In our experiments, the complex medium consists in a series of pillars that are designed in a regular lattice using soft lithography microfabrication. The cells are then introduced in the lattice, and their trajectories within the pillars are tracked and analyzed.  The effect of the complex medium on the swimming behaviour of microswimmers is analyzed through the measure of relevant statistical observables. In particular, the mean correlation time of direction and the effective diffusion coefficient are shown to decrease when increasing the density of pillars. This provides some bases of understanding for active matter in complex environments.
\end{abstract}
\maketitle

%\linenumbers
%\modulolinenumbers[10] %Cada cuanto el conteo de las líneas

\section{Introduction}
Self-propelled particles represent an out-of-equilibrium system of great interest for a large community of physicists \cite{marchetti2013}. The dynamics of most microswimmers, natural or artificial, perform a ``run and tumble''  dynamics of swimming \cite{berg1993}. This terminology, initially dedicated to \textit{E-coli} bacteria, describes an alternation of directed motion at a given velocity - the runs - and reorientation of the direction - the tumbles. Other dynamics of swimming consist in Active Brownian particles where the direction angle changes continuously in a diffusive manner \cite{cates2013active}. These modes of swimming have been shown to be crucial in the search of chemicals or nutrients \cite{mitchell2002}. Depending on  systems, the decorrelation of direction emerges from different mechanisms. In bacteria, tumbles are due to the unbundling of flagella \cite{berg1972chemotaxis}, in the microalga \textit{Chlamydomonas Reinhardtii}, tumbles have been shown to be due to asynchronous periods of beating \cite{polin2009chlamydomonas}. In artificially built microswimmers such as Janus particles, thermal rotational Brownian motion is usually responsi
 ble for the randomisation of orientation \cite{Howse2007,Palacci2010}. Most of the time, the swimming dynamics can be fairly described as a persistent random walk. 

Hence, considering large enough time scales, microswimmers explore their environment in a diffusive-like manner. A nonequilibrium statistical physics framework can then be built in order to deeper understand the behavior of active matter \cite{ramaswamy2010}.  To predict active matter behaviour in realistic conditions such as crowded living tissues, suspensions of cells or porous media, there is a major need to understand the interaction of self-propelled particles with a complex environment \cite{Bechinger2016}. 

In this work, we quantify experimentally the swimming dynamics of a natural microswimmer -- \textit{Chlamydomonas Reinhardtii} -- within a micropatterned environment. We show that the effective diffusivity of microswimmers is hindered by the presence of obstacles, and that the distribution of swimming directions is no longer isotropic.
%Moreover, an analogy with diffusion processes through porous media allows us to rationalize the results in terms of effective diffusivity.
In addition, a theoretical modeling in terms of an effective anistropic scattering medium allows us to relate the anisotropy of swimming directions and the decrease of the effective diffusion coefficient.

\section{Materials and Methods}
The green microalga \textit{Chlamydomonas Reinhardtii} (CR) is a biflagellated photosynthetic cell of about 10 $\mu$m diameter \cite{harris2009chlamydomonas}. Cells are grown under a 14h/10h light/dark cycle at $\unit{22}{\celsius}$ and are harvested in the middle of the exponential growth phase. This microalga propels itself in a break-stroke-type swimming using its two front flagella. 
% Therefore, the persistence or (\textit{Run}) length is then $L_0=v_0 t_c=\unit{180}{\micro\meter}$.

\begin{figure}[!ht]
\centering
  \includegraphics[width=\columnwidth]{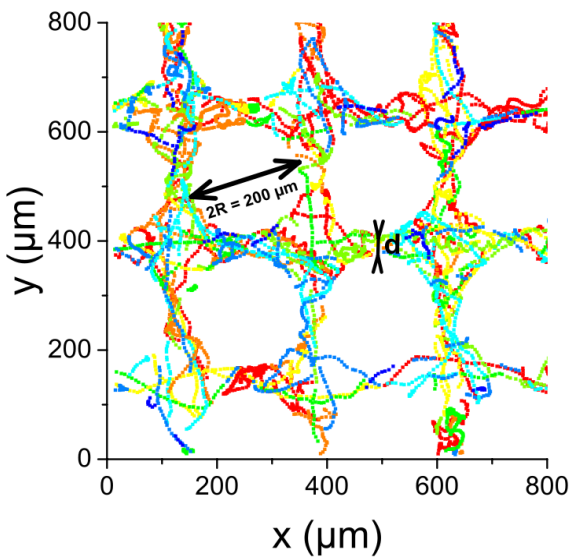}  			       
  \caption{Map of trajectories tracked over 10 seconds within a square lattice of pillars with $d=\unit{50}{\micro\meter}$ at a time interval of $1/15$ s. The color is mapped on the particle index.}  \label{map}
\end{figure}

CR suspensions are used with no further preparation. Suspensions are dilute enough with a volume fraction of about $0.05 \%$, so that hydrodynamic interactions among the particles are negligible.  The cells are then introduced within a complex medium composed of a square lattice of 200 $\mu$m-diameter pillars regularly spaced by a surface-to-surface minimal inter-pillar distance $d$ (figure \ref{map}). The distance $d$ between the surfaces of the pillars ranges from $\unit{20}{\micro\meter}$ to $\unit{50}{\micro\meter}$ with a $\unit{10}{\micro\meter}$ increment, and from $\unit{50}{\micro\meter}$ to $\unit{370}{\micro\meter}$ with a $\unit{40}{\micro\meter}$ increment. This represents a porosity $\delta=1-\pi R^2/(d+2R)^2$ ranging from $0.35$ to $0.90$.  Pillars are made of transparent PDMS using soft lithography processes \cite{qin2010soft}.  Their diameter is kept constant  to $\unit{200}{\micro\meter}$. This is of the same order as the persistence length of the swimming dynamics 
 of the cells ($\sim 180 \mu$m). The height of the pillars is $\unit{70}{\micro\meter}$, which represents about 7 cell diameters. Our control parameter is $d$, which controls the density of the complex medium that the cells experience.

The observations are made by means of bright field microscopy. The chamber is  observed under an inverted microscope (Olympus IX71) coupled to a CMOS camera (Imaging Source)  used at a frame rate of 15 frames per second. A low magnification objective ($\times$1.25) provides a wide field of view of $\unit{800}{\micro\meter}$ $\times$ $\unit{800}{\micro\meter}$  as well as a large depth of field. The sample is enclosed in an occulting box with two red filtered windows for visualisation. The red filters prevent any parasite light that could trigger phototaxis (\textit{i.e }a biased swimming toward a light source) \cite{harris2009chlamydomonas, garciaPRL2013}. 

Particle tracking is performed using Trackpy \cite{trackpy}, a Python library based on Crocker and Grier's algorithm \cite{crocker}. Relevant quantities such as the mean square displacements (MSD) and the correlation functions of directions are then extracted from an ensemble average performed over long lasting movies (6 \minute).

Figure \ref{map} shows the typical geometry and a set of trajectories of cells measured over 10 seconds for a given inter-pillar distance of 50 micrometers ($d=\unit{50}{\micro\meter}$) at a time interval of $1/15$ \second.

\section{Results}

\subsection{Anisotropy}
\begin{figure}[!ht]
\centering
  \includegraphics[width=\columnwidth]{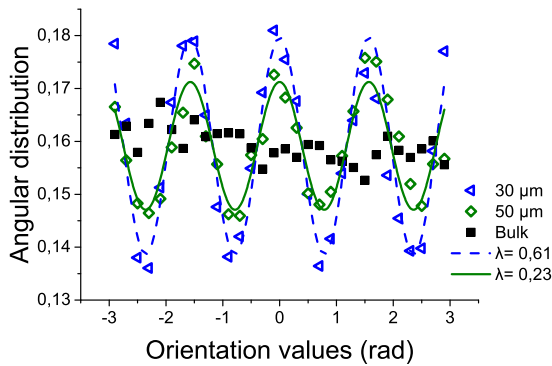}  			       
  \caption{Orientation distributions of Chlamydomonas \textit{runs} for different interpillar distances $d$. To do so, the orientation is measured over $\unit{1}{\second}$. These distributions are fitted with Eq.~(\ref{eq:Ptheta:theo}), assuming $\lambda_0=\lambda_4=\lambda$.}
\label{Orientation_over1s}
\end{figure}

The first noticeable effect of the lattice of obstacles onto the cells is an anisotropy of their swimming directions. The squared lattice of pillars constrains the trajectories of microswimmers to a set of privileged $(x,y)$ directions as shown from the orientation distribution plotted in figure \ref{Orientation_over1s}. Here, we define the orientations as the mean orientation of the trajectory over $1$ s. While orientations of microswimmers are isotropically distributed in a free medium ($d \to \infty$), the distributions show peaks around $\pi/2 (\mathrm{mod}$ $\pi/2)$ when cells are placed within the complex medium. This clearly demonstrates the privileged directions taken by microswimmers. This reflects that, most of the time, the pillars orient the swimming along corridors between pillars. This effect becomes more and more pronounced as $d$ is decreased.

To try to better understand these results, we introduce a relatively simple theoretical model consisting of an active Brownian particle immersed in an effective anisotropic scattering medium.
The active Brownian particle is characterized by its position ${\bf r}$ (in 2D) and an angle $\theta$ defining its direction of motion. The particle moves at a constant speed $v_0$. 
In the absence of obstacles, the angle $\theta$ has a purely diffusive dynamics:
\be \label{eq:def:ABP}
\dot{\bf r} = v_0 {\bf e}(\theta) \,, \quad
\dot{\theta} = \xi(t)
\ee
where $\xi(t)$ is a white noise satisfying $\langle \xi(t) \rangle = 0$ and
\be \label{eq:angular:noise}
\langle \xi(t)\xi(t') \rangle = 2D_R \, \delta(t-t') \,.
\ee
The angular diffusion coefficient is related to the persistence time $\tau$ by
$\tau=1/D_R$.
To make the full problem tractable, the lattice of pillars is modeled as an effective anisotropic scattering medium, with a probability rate
\be \label{eq:Ltheta0}
\lambda(\theta) = \lambda_0 - \lambda_4 \cos(4\theta) \,.
\ee
The two parameters $\lambda_0$ and $\lambda_4$ are constrained by $|\lambda_4| \le \lambda_0$.
After scattering, the new angle $\theta'$ is randomly chosen from a uniform distribution. The model thus boils down to a combination of the active Brownian particle and the run-and-tumble model, with here an anistropic tumbling rate.
Technical details are reported in Appendix~\ref{appendix}. Using some standard approximation techniques, we eventually obtain the spatially averaged probability distribution $\overline{P}(\theta)$ of swimming directions $\theta$
\be \label{eq:Ptheta:theo}
\overline{P}(\theta) = \frac{1}{2\pi} \left( 1+\frac{\lambda_4}{16D_R+\lambda_0} \, \cos(4\theta) \right).
\ee
We use this form to fit the experimental data, under the assumption $\lambda_0=\lambda_4=\lambda$ (which means that particles can travel freely when their direction is aligned either with the $x$ or $y$ axis). This fitting procedure thus allows us to determine the experimental values of $\lambda$.

%\subsection{Geometrical characteristic of the scattering medium}

%\begin{figure}
%\centering
%  \includegraphics[width=\columnwidth]{Dr_lamda0}  			       
%  \caption{$D_{R}/\lambda$ is plotted here as a function of $d/L_{0}$ from the  obtained $\lambda$ values from fitting the angular distributions of fig\ref{Orientation_over1s}. }  \label{lambda}
%\end{figure}

\subsection{Mean square displacements and correlations}  
\begin{figure*}
  %\centering
  \begin{center}
    \begin{tabular}{ll}
    a&b\\
\includegraphics[height=6 cm]{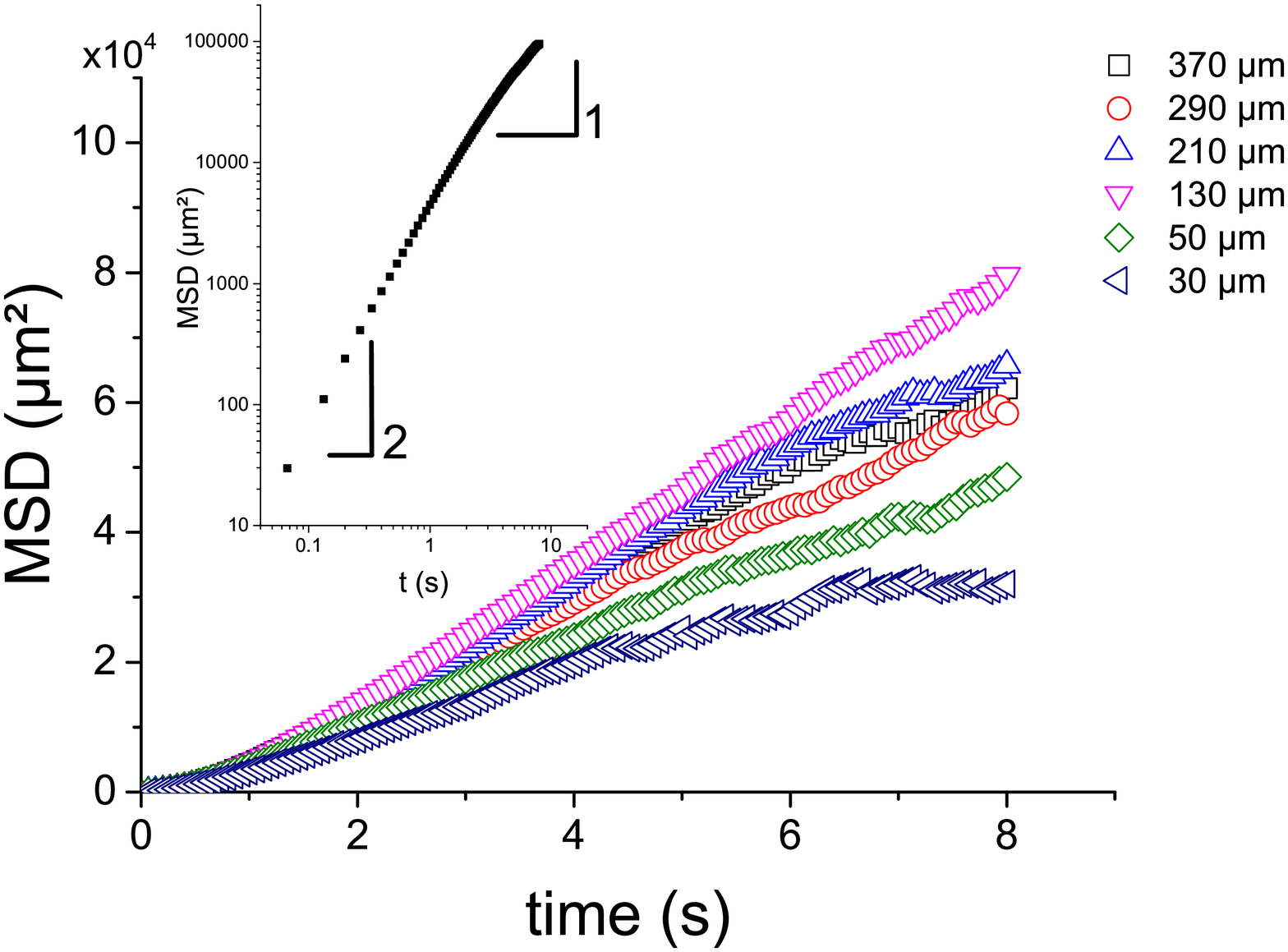}&\includegraphics[height=6 cm]{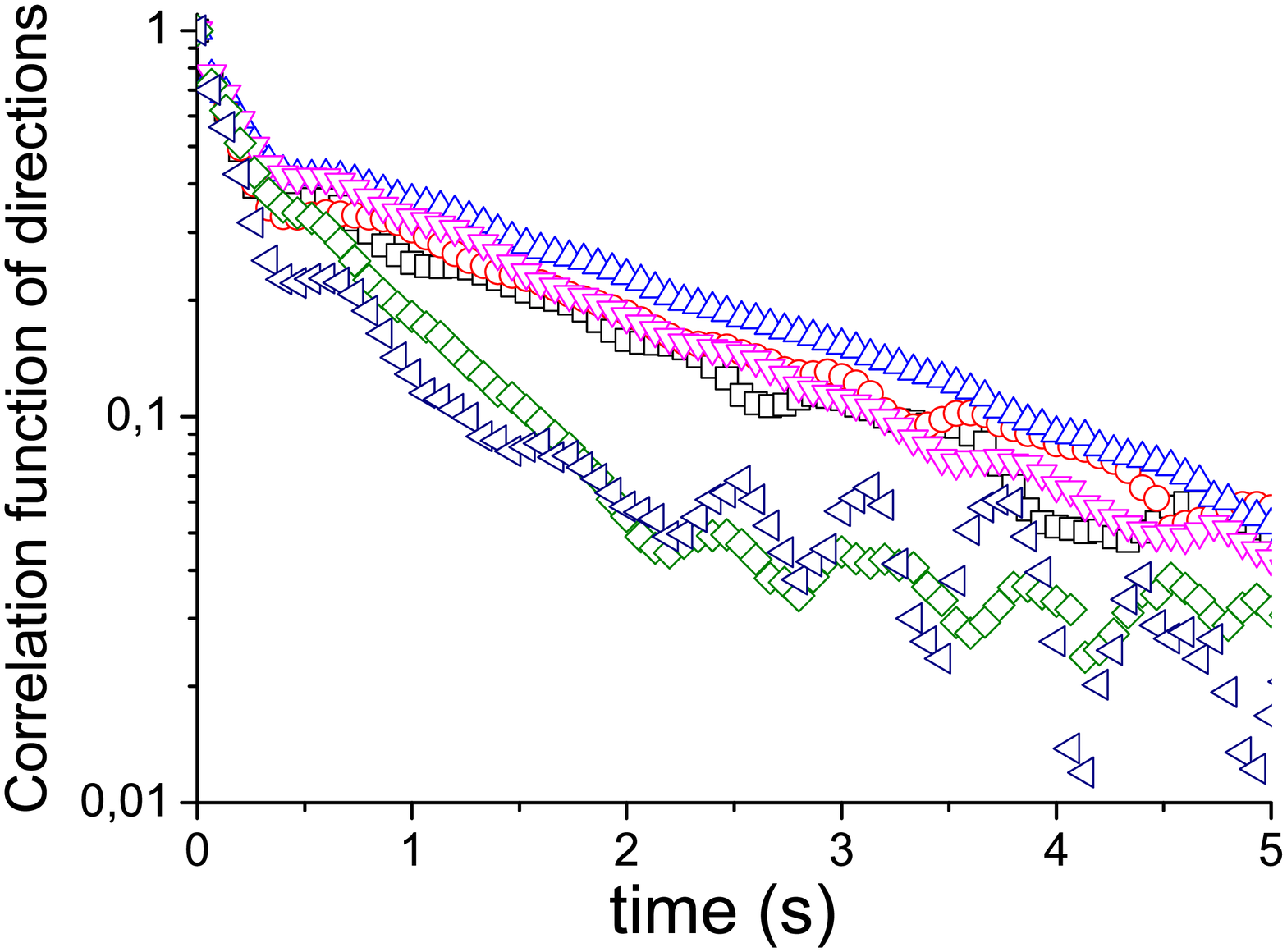}\\
    \end{tabular}
 \end{center}
\caption{a. Mean square displacements for several interpillar distances $d$.  The inset shows the two limiting cases: a slope of $2$ in $log-log$ representation followed by a linear regime at longer timescales. b. Correlation function of direction as a function of time for different values of $d$%Particularly, we found for the swimming in bulk, the persitence length $L_\text{0}=L_\text{eff}(d\to\infty)=\unit{170}{\micro\meter}$ and the correlation time $t_\text{0}=t_\text{eff}(d\to\infty)=\unit{2.6}{\second}$.
 .}\label{msd}
\end{figure*}

From the measured trajectories, we evaluate and plot in figure \ref{msd}-a the mean square displacement (MSD) $\langle\mathbf{r}^2(t)\rangle$ for different values of $d$ ranging from $20$ to $370$ $\mu$m. In a free medium (\textit{i.e.} without pillars), the swimming of CR has been shown to be well characterized by a persistent random walk \cite{polin2009chlamydomonas,Berti,Codling2008} in absence of tropism. Hence, the behaviour of microswimmers can be modelled as a ballistic motion at short timescales (below $ \sim 1$ s) and a diffusive-like one at longer timescales. Here, we assume  that the MSD in the presence of obstacles can still be described by a persistent random walk and we fit the curves in figure \ref{msd} with the following semi-empirical equation:

\begin{equation}
\langle \mathbf{r}^2(t) \rangle = 4 D_{\text{eff}} t - 2 D_{\text{eff}} t_{\text{eff}} \left[ 1-\exp{\left(\frac{-2t}{t_{\text{eff}}}\right)} \right]
\label{eq_msd}
\end{equation}
where $t_{\text{eff}}$ is the effective correlation time and $L_{\text{eff}}$ the effective persistence length of the swimming. In a free medium ($d \to \infty$), we denote by $t_{\text{0}}$ and $L_{\text{0}}$ the correlation time and persistence length respectively. Experimental measurements give $t_{\text{0}} = 2.7 \second $ and $L_{\text{0}}= 180 \micro\meter$

%\begin{figure}
%\centering
%\includegraphics[width=\columnwidth]{msds_inset_SR.eps}
%\caption{Mean square displacements fitted with the equation \ref{eq_msd} for several interpillar distance $d$. Particularly, we found for the swimming in bulk, the persitence length $L_\text{0}=L_\text{eff}(d\to\infty)=\unit{170}{\micro\meter}$ and the correlation time $t_\text{0}=t_\text{eff}(d\to\infty)=\unit{2.6}{\second}$.}\label{msd}
%\end{figure}

In addition, we characterize independently the persistence time by measuring the  correlation
function of direction defined as $$C(t) = \langle{\mathbf{k}}(t_{\text{0}}) \cdot {\mathbf{k}}(t_{\text{0}}+t)\rangle,$$
where  $\langle \dots \rangle$ denotes an average over time $t_0$ and over all
tracked trajectories and $\mathbf{k}$ a unit vector along the trajectory (Figure \ref{msd}-b). Correlations with infinite decay time
($C(t)=1$ for all $t>0$) correspond to swimming directions preserved over
arbitrarily long times characteristic of a purely ballistic regime; in contrast, a
zero life-time ($C(t)=0$ for all $t>0$) corresponds to the standard random walk
behaviour (analogous to Brownian motion). The measured correlation functions show two characteristic times: the first one corresponds to an helical shape \footnote{not discussed here as this is a 3d feature of the trajectories and our microscopy analysis is a 2d study} of the trajectory and the second one $t_{\text{eff}}$ represents the mean time of persistence over which the swimming direction is preserved. The extracted characteristic time $t_{\text{eff}}$ allows one to constraint the fitting procedure of equation (\ref{eq_msd}) and to evaluate $L_{\text{eff}}$ for different values of $d$.

\subsection{Diffusive regime}The long timescales dynamics can be then described by a diffusive-like behaviour. From the measured MSD and the correlation function of direction, we similarly obtain the effective diffusivity $D_{\text{eff}}$ as a function of $d/L_{\text{0}}$.

%The diffusivity increases with $d$ until reaching a plateau whose numerical value is the bulk diffusivity $D_0$. To analyze this regime, we draw an analogy with classical diffusion processes in complex media. Here we do not take into account long range interaction such as hydrodynamic interaction between diffusing objects and the obstacles but only excluded volume considerations. In particular, one can evaluate the obstruction factor, also known as the tortuosity which defines the ratio between the large scale diffusivity $D_{\text{eff}}$ and $D_{\text{0}}$ the bulk solvent diffusion. The obstruction factor depends on the geometrical properties of the obstacles such as shape and volume fraction. The authors of \cite[]{Johannesson1996} propose a numerical calculation of the obstruction factor for infinite cylinders ordered in square lattices which reads (SHOULD WE KEEP THIS EQUATION?):
%
%\begin{equation}
%\frac{D_{\text{eff}}}{D_0} = \frac{1}{1-\phi}\left(1-\frac{2\phi}{1+\phi-0.3058\phi^4}\right), 
% \label{eq_johanesson}
%\end{equation} 
%where the pillars density is related to the inter-pillar distance $d$ and the radius of the pillars as $\phi=\pi R^2/(d+2R)^2$. Free medium corresponds to $\phi = 0$ and $d \to \infty$
%

%Figure \ref{Deff_D0} shows the experimentally measured effective diffusivity of microswimmers within the lattice compared to the numerical expression of equation \ref{eq_johanesson}. The  good agreement suggests that again pure excluded volume might be sufficient to account for the observed features. 

\begin{figure}
\centering
  \includegraphics[width=\columnwidth]{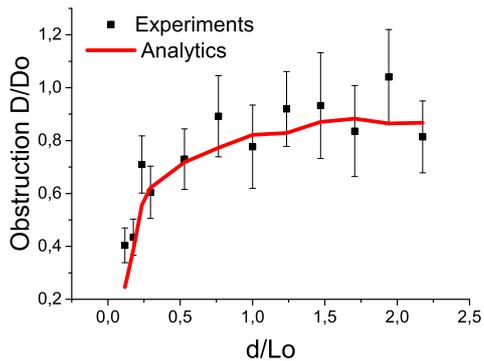}  			       
  \caption{Normalized effective diffusion coefficient $D_{\text{eff}}$ as a function of the interpillar distance $d$. The bulk diffusion, used for normalization, is $D_0=\unit{3000}{\micro\meter^{2}\per\second}$. The continuous line represents the prediction of Eq.~(\ref{eq:DD0:theo}), with $\lambda$ fitted from Eq.~(\ref{eq:Ptheta:theo}).}
\label{Deff_D0}
\end{figure}

Figure~\ref{Deff_D0} shows the experimentally measured effective diffusion coefficient $D_{\text{eff}}$ of microswimmers within the lattice, normalized by the bulk diffusion coefficient $D_0$. These results are compared with the theoretical prediction (see Appendix~\ref{appendix})
\be \label{eq:DD0:theo}
\frac{D}{D_0} = 
\frac{1+\frac{\tilde{\lambda}_4^2}{4(9+\tilde{\lambda}_0)(16+\tilde{\lambda}_0)}}
{(1+\tilde{\lambda_0}) \left(1-\frac{\tilde{\lambda}_4^2}{4(1+\tilde{\lambda}_0)(9+\tilde{\lambda}_0)}\right)}
\ee
where the dimensionless parameters
\be
\tilde{\lambda}_0 = \frac{\lambda_0}{D_R} \,, \quad
\tilde{\lambda}_4 = \frac{\lambda_4}{D_R}
\ee
have been defined. Here, as before we have assumed that both parameters are equal,
and taken the value of $\lambda_0=\lambda_4=\lambda$ obtained from the fit of the distribution of swimming directions (fig. \ref{Orientation_over1s}). This assumption might break down at low interpillar distances. The value of the angular diffusion coefficient corresponding to $1/t_{0}$ is taken as $D_{R} = \unit{0.37}{\second}^{-1} $.

\section{Conclusion}
In this paper, we show that the diffusivity of puller-type microswimmers (here \textit{Chlamydomonas Reinhardtii}) is strongly affected when embedded in a complex medium (here a pillar lattice). We show that geometrical constraints are sufficient to provide a good understanding of the diffusivity as a function of the pillar density. Hydrodynamic interactions can be ignored, at least at low volume fractions of microswimmers. This seems to favour the hypothesis that only steric interactions drive the coupling between pullers and walls \cite{Kantsler2013} rather than the hydrodynamic hypothesis \cite{spagnolie2015geometric, Contino2015, Mirzakhanloo2018}. 

This paves the way to future studies on behaviours of microswimmers hindered by complex geometrical environments.

%%%%%%%%%%%%%%%%%%%%%%%%%%%%%%%%%%%%%%%%%%%%%%%%%%%%%%%
%%%%%%%%%%%%%%%%%%%%%%%%%%%%%%%%%%%%%%%%%%%%%%%%%%%%%%%
%%%%%%%%%%%%%%%%%%%%%%%%%%%%%%%%%%%%%%%%%%%%%%%%%%%%%%%

\appendix

\section{Derivation of an effective diffusive description}
\label{appendix}

In this appendix, we provide an approximate statistical description of the effective large scale diffusion of swimmers, under some simplifying assumptions. The spatially averaged angular distribution of swimmers is also obtained.

We consider a set of active Brownian particles moving in a crowded environment.
Assuming that particles do not interact with each other, we can focus on the description of a single particle. 
The particle is characterized by its position ${\bf r}$ (in 2D) and an angle $\theta$ defining its direction of motion. The particle moves at a constant speed $v_0$. 
In the absence of obstacles, the angle $\theta$ has a purely diffusive dynamics, and the model is described by Eqs.~(\ref{eq:def:ABP}) and (\ref{eq:angular:noise}).
To introduce the pillars in a simplified way so that the problem remains analytically tractable, we use an effective medium approach. As a first approximation, the effect of pillars is to generate random changes in the direction of motion of the swimmers. To simplify the problem, we neglect spatial correlations and simply retain as a key ingredient the 4-fold anisotropy resulting from the lattice of pillars.
We then describe the crowded environment by a stochastic probability $\lambda(\theta)$ to change direction after collision with a pillar.
As for run-and-tumble particles, we assume that the new direction $\theta'$ is chosen in a uniform way. Doing so, we neglect the probable anticorrelation between $\theta$ and $\theta'$ (one expects that the swimmer is more likely to go backward after a collision with a pillar, but this effect is probably not very strong).

The anisotropy of the medium is kept in the description through the $\theta$ dependence of the `tumbling' rate $\lambda(\theta)$, and we choose as the simplest description to keep only the zeroth and fourth angular mode, leading to
\be \label{eq:Ltheta}
\lambda(\theta) = \lambda_0 - \lambda_4 \cos(4\theta) \,.
\ee
The positivity of $\lambda(\theta)$ implies that $\lambda_0 >0$ and $|\lambda_4| \le \lambda_0$.
In the experiment, the arrangement of pillars corresponds to an effective medium that is less crowded along the $x$ and $y$ axes than along the diagonal directions. We have chosen the sign convention in Eq.~(\ref{eq:Ltheta}) so that the experimental situation corresponds to $\lambda_4 >0$.

%To make an explicit connection with the pillar geometry, the probability rate $\lambda(\theta)$ may be expressed as
%\be
%\lambda(\theta) = \frac{v_0}{L(\theta)}
%\ee
%where $L(\theta)$ is the average length to the next obstacle when moving in the direction $\theta$, starting from a random location.
%Eq.~(\ref{eq:Ltheta}) is then obtained by an angular Fourier expansion, keeping only leading order terms.

\subsubsection*{Effective diffusion equation}

The probability density $P({\bf r},\theta)$ to find the swimmer at position ${\bf r}$ with velocity angle $\theta$ obeys the following dynamics:
\bea \nonumber
&& \partial_t P({\bf r},\theta) + \nabla \cdot [v_0 {\bf e}(\theta) P({\bf r},\theta)] = D_R \partial_{\theta}^2 P({\bf r},\theta) \\ 
&& \qquad -\lambda(\theta) P({\bf r},\theta)
+ \frac{1}{2\pi} \int d\theta' \, \lambda(\theta') P({\bf r},\theta')
\label{eq:dyn:Prtheta}
\eea
We define the angular Fourier expansion of the distribution $P({\bf r},\theta)$,
\bea
P({\bf r},\theta) &=& \frac{1}{2\pi} \sum_{k=-\infty}^{\infty} f_k({\bf r})
\, e^{-ik\theta}\\
f_k({\bf r}) &=& \int_{-\pi}^{\pi} d\theta \, P({\bf r},\theta) \, e^{ik\theta}
\eea
Note that $\rho({\bf r}) \equiv f_0({\bf r})$ is simply the density field.
Expanding Eq.~(\ref{eq:dyn:Prtheta}) in angular Fourier modes, we get
\bea \nonumber
\partial_t f_k &+& \frac{v_0}{2} \left( \hat\triangledown f_{k-1}
+ \hat\triangledown^* f_{k+1} \right) = - D_R k^2 f_k  \\ \nonumber
&& \quad - \lambda_0 f_k + \frac{\lambda_4}{2} \left( f_{k+4} + f_{k-4} \right) \\
&& \quad + \delta_{k,0} \left[ \lambda_0 \rho - \frac{\lambda_4}{2}
\left( f_4 + f_4^* \right) \right]
\label{eq:Fourier}
\eea
where the star denotes the complex conjugate, and $\delta_{k,0}=1$ for $k=0$ and $0$ otherwise. The notations $\hat\triangledown$ and $\hat\triangledown^*$ denote the complex differential operator
\be
\hat\triangledown = \partial_x + i \partial_y \,, \quad
\hat\triangledown^* = \partial_x - i \partial_y \,,
\ee
with $i^2=-1$.
For $k=0$, Eq.~(\ref{eq:Fourier}) simply yields a continuity equation
\be \label{eq:continuity}
\partial_t \rho + \frac{v_0}{2} {\rm Re}(\hat\triangledown^* f_1) = 0
\ee
which is equivalent to the standard continuity equation
\be
\partial_t \rho + \nabla \cdot (\rho {\bf v}) = 0 \,,
\ee
where ${\bf v}$ is the hydrodynamic velocity field,
with $\rho {\bf v}=v_0({\rm Re} f_1,{\rm Im} f_1)$.
The goal of the following study is to obtain a closed expression of the field $f_1$ in terms of the density field $\rho$ and its derivatives, thus turning Eq.~(\ref{eq:continuity}) into a closed differential equation on the field $\rho$.
To express $f_1$ as a function of $\rho$ and its derivatives, we need to rely on Eq.~(\ref{eq:Fourier}), which corresponds to an infinite hierarchy of coupled equations.
To make the problem tractable, we need to truncate this hierarchy. Because of the 4-fold symmetry of the problem, we need to keep angular modes at least up to $f_4$. In what follows, we neglect all modes $f_k$ with $|k| >4$.
For $k \neq 0$, the dynamics of $f_k$ involves a relaxation term
$-(D_Rk^2+\lambda_0)f_k$, while the density field does not have such a relaxation dynamics. Hence the density field $\rho$ is a `slow' variable, while the fields $f_{k\neq 0}$ are `fast' variables.
As a result, on time scales larger than $1/D_R$, the time derivatives $\partial_t f_k$ can be neglected for $k \neq 0$.
We end up with the following set of equations
\bea
\label{eq:f1}
f_1 &=& -a_1 \left( \hat\triangledown \rho + \hat\triangledown^* f_2 \right) + b_1 f_3^* \\
\label{eq:f2}
f_2 &=& -a_2 \left( \hat\triangledown f_1 + \hat\triangledown^* f_3 \right) + b_2 f_2^* \\
\label{eq:f3}
f_3 &=& -a_3 \left( \hat\triangledown f_2 + \hat\triangledown^* f_4 \right) + b_3 f_1^* \\
\label{eq:f4}
f_4 &=& -a_4 \hat\triangledown f_3 + b_4 \rho
\eea
where the coefficients are given by
\be
a_k = \frac{v_0}{2(k^2 D_R + \lambda_0)} \,, \quad
b_k = \frac{\lambda_4}{2(k^2 D_R + \lambda_0)} \,.
\ee
We wish to determine an effective diffusion equation from Eq.~(\ref{eq:continuity}), and thus we need to express $f_1$ as a function of $\rho$ up to gradient order.
Eq.~(\ref{eq:f1}) provides an expression for $f_1$ in terms of the fields $\rho$, $f_2$ and $f_3$. We need to determine $f_2$ to zeroth order in gradient, and $f_3$ to first order in gradient.
Eq.~(\ref{eq:f2}) shows that at zeroth order in gradient, $f_2=0$.
Then combining Eqs.~(\ref{eq:f1}), (\ref{eq:f3}) and (\ref{eq:f4}), we get
at first order in gradient
\be
f_3 = - \frac{a_3 b_4 + b_3 a_1}{1-b_1b_3} \, \hat\triangledown^* \rho
\ee
and thus from Eq.~(\ref{eq:f1}),
\be
f_1 = -\frac{a_1 + a_3 b_1 b_4}{1-b_1b_3} \, \hat\triangledown \rho
\ee
(again to first order in gradients).
We thus finally obtain, using Eq.~(\ref{eq:continuity}), the diffusion equation
\be
\partial_t \rho = D \Delta \rho
\ee
with a diffusion coefficient
\be
D = \frac{v_0(a_1 + a_3 b_1 b_4)}{1-b_1b_3}
\ee
In the absence of obstacles, $D$ reduces to the well-known diffusion coefficient of active Brownian particles,
\be
D_0 = \frac{v_0^2}{2D_R}
\ee
It is convenient to define the dimensionless parameters
\be
\tilde{\lambda}_0 = \frac{\lambda_0}{D_R} \,, \quad
\tilde{\lambda}_4 = \frac{\lambda_4}{D_R}
\ee
With these notations, the ratio $D/D_0$ can be explicitly expressed as
\be
\frac{D}{D_0} = 
\frac{1+\frac{\tilde{\lambda}_4^2}{4(9+\tilde{\lambda}_0)(16+\tilde{\lambda}_0)}}
{(1+\tilde{\lambda_0}) \left(1-\frac{\tilde{\lambda}_4^2}{4(1+\tilde{\lambda}_0)(9+\tilde{\lambda}_0)}\right)}
\ee
Finally, the mean square displacement is given by
\be
\langle {\bf r}^2(t) \rangle = 4Dt
\ee

\subsubsection*{Spatially averaged angular distribution}

To see if the dynamics of the active Brownian particle keeps track, on large scale, of the anisotropy of the medium, we compute the spatially integrated angular distribution
\be
\overline{P}(\theta) = \int d{\bf r} \, P({\bf r},\theta)
\ee
Keeping as above only angular Fourier modes up to $|k|=4$, we have
\be
\overline{P}(\theta) = \frac{1}{2\pi} \sum_{k=-4}^{4} \overline{f}_k({\bf r})
\, e^{-ik\theta} \,.
\ee
Given that the integral over space of space derivative terms is equal to zero, we get from Eqs.~(\ref{eq:f1}), (\ref{eq:f2}) and (\ref{eq:f3}),
\be
\overline{f}_1 = b_1\overline{f}_3^* \,, \quad
\overline{f}_2 = b_2\overline{f}_2^* \,, \quad
\overline{f}_3 = b_3\overline{f}_1^*
\ee
from which it is easy to show that 
$\overline{f}_1 = \overline{f}_2 = \overline{f}_3=0$.
In addition, we have from Eq.~(\ref{eq:f4}) that $\overline{f}_4 = b_4 \overline{\rho}=b_4$, 
because $\int d{\bf r} \, \rho({\bf r}) = 1$ since we have considered a single particle.
We eventually get
\be
\overline{P}(\theta) = \frac{1}{2\pi} \left( 1+\frac{\lambda_4}{16D_R+\lambda_0} \, \cos(4\theta) \right).
\ee

%biblio

%%%%%%%%%%%%%


%\bibliography{bibi}

\begin{thebibliography}{22}%
\makeatletter
\providecommand \@ifxundefined [1]{%
 \@ifx{#1\undefined}
}%
\providecommand \@ifnum [1]{%
 \ifnum #1\expandafter \@firstoftwo
 \else \expandafter \@secondoftwo
 \fi
}%
\providecommand \@ifx [1]{%
 \ifx #1\expandafter \@firstoftwo
 \else \expandafter \@secondoftwo
 \fi
}%
\providecommand \natexlab [1]{#1}%
\providecommand \enquote  [1]{``#1''}%
\providecommand \bibnamefont  [1]{#1}%
\providecommand \bibfnamefont [1]{#1}%
\providecommand \citenamefont [1]{#1}%
\providecommand \href@noop [0]{\@secondoftwo}%
\providecommand \href [0]{\begingroup \@sanitize@url \@href}%
\providecommand \@href[1]{\@@startlink{#1}\@@href}%
\providecommand \@@href[1]{\endgroup#1\@@endlink}%
\providecommand \@sanitize@url [0]{\catcode `\\12\catcode `\$12\catcode
  `\&12\catcode `\#12\catcode `\^12\catcode `\_12\catcode `\%12\relax}%
\providecommand \@@startlink[1]{}%
\providecommand \@@endlink[0]{}%
\providecommand \url  [0]{\begingroup\@sanitize@url \@url }%
\providecommand \@url [1]{\endgroup\@href {#1}{\urlprefix }}%
\providecommand \urlprefix  [0]{URL }%
\providecommand \Eprint [0]{\href }%
\providecommand \doibase [0]{http://dx.doi.org/}%
\providecommand \selectlanguage [0]{\@gobble}%
\providecommand \bibinfo  [0]{\@secondoftwo}%
\providecommand \bibfield  [0]{\@secondoftwo}%
\providecommand \translation [1]{[#1]}%
\providecommand \BibitemOpen [0]{}%
\providecommand \bibitemStop [0]{}%
\providecommand \bibitemNoStop [0]{.\EOS\space}%
\providecommand \EOS [0]{\spacefactor3000\relax}%
\providecommand \BibitemShut  [1]{\csname bibitem#1\endcsname}%
\let\auto@bib@innerbib\@empty
%</preamble>
\bibitem [{\citenamefont {Marchetti}\ \emph {et~al.}(2013)\citenamefont
  {Marchetti}, \citenamefont {Joanny}, \citenamefont {Ramaswamy}, \citenamefont
  {Liverpool}, \citenamefont {Prost}, \citenamefont {Rao},\ and\ \citenamefont
  {Simha}}]{marchetti2013}%
  \BibitemOpen
  \bibfield  {author} {\bibinfo {author} {\bibfnamefont {M.~C.}\ \bibnamefont
  {Marchetti}}, \bibinfo {author} {\bibfnamefont {J.~F.}\ \bibnamefont
  {Joanny}}, \bibinfo {author} {\bibfnamefont {S.}~\bibnamefont {Ramaswamy}},
  \bibinfo {author} {\bibfnamefont {T.~B.}\ \bibnamefont {Liverpool}}, \bibinfo
  {author} {\bibfnamefont {J.}~\bibnamefont {Prost}}, \bibinfo {author}
  {\bibfnamefont {M.}~\bibnamefont {Rao}}, \ and\ \bibinfo {author}
  {\bibfnamefont {R.~A.}\ \bibnamefont {Simha}},\ }\href {\doibase
  10.1103/RevModPhys.85.1143} {\bibfield  {journal} {\bibinfo  {journal} {Rev.
  Mod. Phys.}\ }\textbf {\bibinfo {volume} {85}},\ \bibinfo {pages} {1143}
  (\bibinfo {year} {2013})}\BibitemShut {NoStop}%
\bibitem [{\citenamefont {Berg}(1993)}]{berg1993}%
  \BibitemOpen
  \bibfield  {author} {\bibinfo {author} {\bibfnamefont {H.~C.}\ \bibnamefont
  {Berg}},\ }\href@noop {} {\emph {\bibinfo {title} {Random walks in
  biology}}}\ (\bibinfo  {publisher} {Princeton University Press},\ \bibinfo
  {year} {1993})\BibitemShut {NoStop}%
\bibitem [{\citenamefont {Cates}\ and\ \citenamefont
  {Tailleur}(2013)}]{cates2013active}%
  \BibitemOpen
  \bibfield  {author} {\bibinfo {author} {\bibfnamefont {M.}~\bibnamefont
  {Cates}}\ and\ \bibinfo {author} {\bibfnamefont {J.}~\bibnamefont
  {Tailleur}},\ }\href@noop {} {\bibfield  {journal} {\bibinfo  {journal} {EPL
  (Europhysics Letters)}\ }\textbf {\bibinfo {volume} {101}},\ \bibinfo {pages}
  {20010} (\bibinfo {year} {2013})}\BibitemShut {NoStop}%
\bibitem [{\citenamefont {Mitchell}(2002)}]{mitchell2002}%
  \BibitemOpen
  \bibfield  {author} {\bibinfo {author} {\bibfnamefont {J.~G.}\ \bibnamefont
  {Mitchell}},\ }\href@noop {} {\bibfield  {journal} {\bibinfo  {journal} {The
  American Naturalist}\ }\textbf {\bibinfo {volume} {160}},\ \bibinfo {pages}
  {727} (\bibinfo {year} {2002})}\BibitemShut {NoStop}%
\bibitem [{\citenamefont {Berg}\ \emph {et~al.}(1972)\citenamefont {Berg},
  \citenamefont {Brown} \emph {et~al.}}]{berg1972chemotaxis}%
  \BibitemOpen
  \bibfield  {author} {\bibinfo {author} {\bibfnamefont {H.~C.}\ \bibnamefont
  {Berg}}, \bibinfo {author} {\bibfnamefont {D.~A.}\ \bibnamefont {Brown}},
  \emph {et~al.},\ }\href@noop {} {\bibfield  {journal} {\bibinfo  {journal}
  {Nature}\ }\textbf {\bibinfo {volume} {239}},\ \bibinfo {pages} {500}
  (\bibinfo {year} {1972})}\BibitemShut {NoStop}%
\bibitem [{\citenamefont {Polin}\ \emph {et~al.}(2009)\citenamefont {Polin},
  \citenamefont {Tuval}, \citenamefont {Drescher}, \citenamefont {Gollub},\
  and\ \citenamefont {Goldstein}}]{polin2009chlamydomonas}%
  \BibitemOpen
  \bibfield  {author} {\bibinfo {author} {\bibfnamefont {M.}~\bibnamefont
  {Polin}}, \bibinfo {author} {\bibfnamefont {I.}~\bibnamefont {Tuval}},
  \bibinfo {author} {\bibfnamefont {K.}~\bibnamefont {Drescher}}, \bibinfo
  {author} {\bibfnamefont {J.~P.}\ \bibnamefont {Gollub}}, \ and\ \bibinfo
  {author} {\bibfnamefont {R.~E.}\ \bibnamefont {Goldstein}},\ }\href {\doibase
  10.1126/science.1172667} {\bibfield  {journal} {\bibinfo  {journal}
  {Science}\ }\textbf {\bibinfo {volume} {325}},\ \bibinfo {pages} {487}
  (\bibinfo {year} {2009})}\BibitemShut {NoStop}%
\bibitem [{\citenamefont {Howse}\ \emph {et~al.}(2007)\citenamefont {Howse},
  \citenamefont {Jones}, \citenamefont {Ryan}, \citenamefont {Gough},
  \citenamefont {Vafabakhsh},\ and\ \citenamefont {Golestanian}}]{Howse2007}%
  \BibitemOpen
  \bibfield  {author} {\bibinfo {author} {\bibfnamefont {J.~R.}\ \bibnamefont
  {Howse}}, \bibinfo {author} {\bibfnamefont {R.~A.~L.}\ \bibnamefont {Jones}},
  \bibinfo {author} {\bibfnamefont {A.~J.}\ \bibnamefont {Ryan}}, \bibinfo
  {author} {\bibfnamefont {T.}~\bibnamefont {Gough}}, \bibinfo {author}
  {\bibfnamefont {R.}~\bibnamefont {Vafabakhsh}}, \ and\ \bibinfo {author}
  {\bibfnamefont {R.}~\bibnamefont {Golestanian}},\ }\href {\doibase
  10.1103/PhysRevLett.99.048102} {\bibfield  {journal} {\bibinfo  {journal}
  {Phys. Rev. Lett.}\ }\textbf {\bibinfo {volume} {99}},\ \bibinfo {pages}
  {048102} (\bibinfo {year} {2007})}\BibitemShut {NoStop}%
\bibitem [{\citenamefont {Palacci}\ \emph {et~al.}(2010)\citenamefont
  {Palacci}, \citenamefont {Cottin-Bizonne}, \citenamefont {Ybert},\ and\
  \citenamefont {Bocquet}}]{Palacci2010}%
  \BibitemOpen
  \bibfield  {author} {\bibinfo {author} {\bibfnamefont {J.}~\bibnamefont
  {Palacci}}, \bibinfo {author} {\bibfnamefont {C.}~\bibnamefont
  {Cottin-Bizonne}}, \bibinfo {author} {\bibfnamefont {C.}~\bibnamefont
  {Ybert}}, \ and\ \bibinfo {author} {\bibfnamefont {L.}~\bibnamefont
  {Bocquet}},\ }\href {\doibase 10.1103/PhysRevLett.105.088304} {\bibfield
  {journal} {\bibinfo  {journal} {Phys. Rev. Lett.}\ }\textbf {\bibinfo
  {volume} {105}},\ \bibinfo {pages} {088304} (\bibinfo {year}
  {2010})}\BibitemShut {NoStop}%
\bibitem [{\citenamefont {Ramaswamy}(2010)}]{ramaswamy2010}%
  \BibitemOpen
  \bibfield  {author} {\bibinfo {author} {\bibfnamefont {S.}~\bibnamefont
  {Ramaswamy}},\ }\href {\doibase 10.1146/annurev-conmatphys-070909-104101}
  {\bibfield  {journal} {\bibinfo  {journal} {Annual Review of Condensed Matter
  Physics}\ }\textbf {\bibinfo {volume} {1}} (\bibinfo {year} {2010}),\
  10.1146/annurev-conmatphys-070909-104101}\BibitemShut {NoStop}%
\bibitem [{\citenamefont {Bechinger}\ \emph {et~al.}(2016)\citenamefont
  {Bechinger}, \citenamefont {Di~Leonardo}, \citenamefont {L\"owen},
  \citenamefont {Reichhardt}, \citenamefont {Volpe},\ and\ \citenamefont
  {Volpe}}]{Bechinger2016}%
  \BibitemOpen
  \bibfield  {author} {\bibinfo {author} {\bibfnamefont {C.}~\bibnamefont
  {Bechinger}}, \bibinfo {author} {\bibfnamefont {R.}~\bibnamefont
  {Di~Leonardo}}, \bibinfo {author} {\bibfnamefont {H.}~\bibnamefont
  {L\"owen}}, \bibinfo {author} {\bibfnamefont {C.}~\bibnamefont {Reichhardt}},
  \bibinfo {author} {\bibfnamefont {G.}~\bibnamefont {Volpe}}, \ and\ \bibinfo
  {author} {\bibfnamefont {G.}~\bibnamefont {Volpe}},\ }\href@noop {}
  {\bibfield  {journal} {\bibinfo  {journal} {Reviews of Modern Physics}\
  }\textbf {\bibinfo {volume} {88}},\ \bibinfo {pages} {045006} (\bibinfo
  {year} {2016})}\BibitemShut {NoStop}%
\bibitem [{\citenamefont {Harris}(2009)}]{harris2009chlamydomonas}%
  \BibitemOpen
  \bibfield  {author} {\bibinfo {author} {\bibfnamefont {E.~H.}\ \bibnamefont
  {Harris}},\ }\href@noop {} {\emph {\bibinfo {title} {The Chlamydomonas
  sourcebook: introduction to Chlamydomonas and its laboratory use}}},\
  Vol.~\bibinfo {volume} {1}\ (\bibinfo  {publisher} {Academic press},\
  \bibinfo {year} {2009})\BibitemShut {NoStop}%
\bibitem [{\citenamefont {Qin}\ \emph {et~al.}(2010)\citenamefont {Qin},
  \citenamefont {Xia},\ and\ \citenamefont {Whitesides}}]{qin2010soft}%
  \BibitemOpen
  \bibfield  {author} {\bibinfo {author} {\bibfnamefont {D.}~\bibnamefont
  {Qin}}, \bibinfo {author} {\bibfnamefont {Y.}~\bibnamefont {Xia}}, \ and\
  \bibinfo {author} {\bibfnamefont {G.~M.}\ \bibnamefont {Whitesides}},\
  }\href@noop {} {\bibfield  {journal} {\bibinfo  {journal} {Nature protocols}\
  }\textbf {\bibinfo {volume} {5}},\ \bibinfo {pages} {491} (\bibinfo {year}
  {2010})}\BibitemShut {NoStop}%
\bibitem [{\citenamefont {Garcia}\ \emph {et~al.}(2013)\citenamefont {Garcia},
  \citenamefont {Rafa{\"\i}},\ and\ \citenamefont {Peyla}}]{garciaPRL2013}%
  \BibitemOpen
  \bibfield  {author} {\bibinfo {author} {\bibfnamefont {X.}~\bibnamefont
  {Garcia}}, \bibinfo {author} {\bibfnamefont {S.}~\bibnamefont {Rafa{\"\i}}},
  \ and\ \bibinfo {author} {\bibfnamefont {P.}~\bibnamefont {Peyla}},\
  }\href@noop {} {\bibfield  {journal} {\bibinfo  {journal} {Physical review
  letters}\ }\textbf {\bibinfo {volume} {110}},\ \bibinfo {pages} {138106}
  (\bibinfo {year} {2013})}\BibitemShut {NoStop}%
\bibitem [{\citenamefont {Allan}\ \emph {et~al.}(2018)\citenamefont {Allan},
  \citenamefont {Caswell}, \citenamefont {Keim},\ and\ \citenamefont {van~der
  Wel}}]{trackpy}%
  \BibitemOpen
  \bibfield  {author} {\bibinfo {author} {\bibfnamefont {D.~B.}\ \bibnamefont
  {Allan}}, \bibinfo {author} {\bibfnamefont {T.}~\bibnamefont {Caswell}},
  \bibinfo {author} {\bibfnamefont {N.~C.}\ \bibnamefont {Keim}}, \ and\
  \bibinfo {author} {\bibfnamefont {C.~M.}\ \bibnamefont {van~der Wel}},\
  }\href {\doibase 10.5281/zenodo.1226458} {\enquote {\bibinfo {title} {Trackpy
  v0.4.1},}\ } (\bibinfo {year} {2018})\BibitemShut {NoStop}%
\bibitem [{\citenamefont {Crocker}\ and\ \citenamefont
  {Grier}(1996)}]{crocker}%
  \BibitemOpen
  \bibfield  {author} {\bibinfo {author} {\bibfnamefont {J.~C.}\ \bibnamefont
  {Crocker}}\ and\ \bibinfo {author} {\bibfnamefont {D.~G.}\ \bibnamefont
  {Grier}},\ }\href@noop {} {\bibfield  {journal} {\bibinfo  {journal} {Journal
  of colloid and interface science}\ }\textbf {\bibinfo {volume} {179}},\
  \bibinfo {pages} {298} (\bibinfo {year} {1996})}\BibitemShut {NoStop}%
\bibitem [{\citenamefont {Garcia}\ \emph {et~al.}(2011)\citenamefont {Garcia},
  \citenamefont {Berti}, \citenamefont {Peyla},\ and\ \citenamefont
  {Rafa\"{\i}}}]{Berti}%
  \BibitemOpen
  \bibfield  {author} {\bibinfo {author} {\bibfnamefont {M.}~\bibnamefont
  {Garcia}}, \bibinfo {author} {\bibfnamefont {S.}~\bibnamefont {Berti}},
  \bibinfo {author} {\bibfnamefont {P.}~\bibnamefont {Peyla}}, \ and\ \bibinfo
  {author} {\bibfnamefont {S.}~\bibnamefont {Rafa\"{\i}}},\ }\href {\doibase
  10.1103/PhysRevE.83.035301} {\bibfield  {journal} {\bibinfo  {journal} {Phys.
  Rev. E}\ }\textbf {\bibinfo {volume} {83}},\ \bibinfo {pages} {035301}
  (\bibinfo {year} {2011})}\BibitemShut {NoStop}%
\bibitem [{\citenamefont {Codling}\ \emph {et~al.}(2008)\citenamefont
  {Codling}, \citenamefont {Plank},\ and\ \citenamefont
  {Benhamou}}]{Codling2008}%
  \BibitemOpen
  \bibfield  {author} {\bibinfo {author} {\bibfnamefont {E.~A.}\ \bibnamefont
  {Codling}}, \bibinfo {author} {\bibfnamefont {M.~J.}\ \bibnamefont {Plank}},
  \ and\ \bibinfo {author} {\bibfnamefont {S.}~\bibnamefont {Benhamou}},\
  }\href@noop {} {\bibfield  {journal} {\bibinfo  {journal} {Journal of the
  Royal Society Interface}\ }\textbf {\bibinfo {volume} {5}},\ \bibinfo {pages}
  {813} (\bibinfo {year} {2008})}\BibitemShut {NoStop}%
\bibitem [{Note1()}]{Note1}%
  \BibitemOpen
  \bibinfo {note} {Not discussed here as this is a 3d feature of the
  trajectories and our microscopy analysis is a 2d study}\BibitemShut {NoStop}%
\bibitem [{\citenamefont {Kantsler}\ \emph {et~al.}(2013)\citenamefont
  {Kantsler}, \citenamefont {Dunkel}, \citenamefont {Polin},\ and\
  \citenamefont {Goldstein}}]{Kantsler2013}%
  \BibitemOpen
  \bibfield  {author} {\bibinfo {author} {\bibfnamefont {V.}~\bibnamefont
  {Kantsler}}, \bibinfo {author} {\bibfnamefont {J.}~\bibnamefont {Dunkel}},
  \bibinfo {author} {\bibfnamefont {M.}~\bibnamefont {Polin}}, \ and\ \bibinfo
  {author} {\bibfnamefont {R.~E.}\ \bibnamefont {Goldstein}},\ }\href {\doibase
  10.1073/pnas.1210548110} {\bibfield  {journal} {\bibinfo  {journal}
  {Proceedings of the National Academy of Sciences}\ }\textbf {\bibinfo
  {volume} {110}},\ \bibinfo {pages} {1187} (\bibinfo {year} {2013})},\ \Eprint
  {http://arxiv.org/abs/http://www.pnas.org/content/110/4/1187.full.pdf}
  {http://www.pnas.org/content/110/4/1187.full.pdf} \BibitemShut {NoStop}%
\bibitem [{\citenamefont {Spagnolie}\ \emph {et~al.}(2015)\citenamefont
  {Spagnolie}, \citenamefont {Moreno-Flores}, \citenamefont {Bartolo},\ and\
  \citenamefont {Lauga}}]{spagnolie2015geometric}%
  \BibitemOpen
  \bibfield  {author} {\bibinfo {author} {\bibfnamefont {S.~E.}\ \bibnamefont
  {Spagnolie}}, \bibinfo {author} {\bibfnamefont {G.~R.}\ \bibnamefont
  {Moreno-Flores}}, \bibinfo {author} {\bibfnamefont {D.}~\bibnamefont
  {Bartolo}}, \ and\ \bibinfo {author} {\bibfnamefont {E.}~\bibnamefont
  {Lauga}},\ }\href@noop {} {\bibfield  {journal} {\bibinfo  {journal} {Soft
  Matter}\ }\textbf {\bibinfo {volume} {11}},\ \bibinfo {pages} {3396}
  (\bibinfo {year} {2015})}\BibitemShut {NoStop}%
\bibitem [{\citenamefont {Contino}\ \emph {et~al.}(2015)\citenamefont
  {Contino}, \citenamefont {Lushi}, \citenamefont {Tuval}, \citenamefont
  {Kantsler},\ and\ \citenamefont {Polin}}]{Contino2015}%
  \BibitemOpen
  \bibfield  {author} {\bibinfo {author} {\bibfnamefont {M.}~\bibnamefont
  {Contino}}, \bibinfo {author} {\bibfnamefont {E.}~\bibnamefont {Lushi}},
  \bibinfo {author} {\bibfnamefont {I.}~\bibnamefont {Tuval}}, \bibinfo
  {author} {\bibfnamefont {V.}~\bibnamefont {Kantsler}}, \ and\ \bibinfo
  {author} {\bibfnamefont {M.}~\bibnamefont {Polin}},\ }\href {\doibase
  10.1103/PhysRevLett.115.258102} {\bibfield  {journal} {\bibinfo  {journal}
  {Physical Review Letters}\ }\textbf {\bibinfo {volume} {115}},\ \bibinfo
  {pages} {1} (\bibinfo {year} {2015})},\ \Eprint
  {http://arxiv.org/abs/1511.00888} {arXiv:1511.00888} \BibitemShut {NoStop}%
\bibitem [{\citenamefont {Mirzakhanloo}\ and\ \citenamefont
  {Alam}(2018)}]{Mirzakhanloo2018}%
  \BibitemOpen
  \bibfield  {author} {\bibinfo {author} {\bibfnamefont {M.}~\bibnamefont
  {Mirzakhanloo}}\ and\ \bibinfo {author} {\bibfnamefont {M.-R.}\ \bibnamefont
  {Alam}},\ }\href {\doibase 10.1103/PhysRevE.98.012603} {\bibfield  {journal}
  {\bibinfo  {journal} {Phys. Rev. E}\ }\textbf {\bibinfo {volume} {98}},\
  \bibinfo {pages} {012603} (\bibinfo {year} {2018})}\BibitemShut {NoStop}%
\end{thebibliography}
%
\end{document}